\documentclass[final,1p,times]{elsarticle}

\usepackage{graphicx}

\usepackage{amssymb}

\journal{Nuclear Physics A}

\begin{document}

\begin{frontmatter}



\long\def\symbolfootnote[#1]#2{\begingroup%
\def\thefootnote{\fnsymbol{footnote}}\footnote[#1]{#2}\endgroup}

\title{Preliminary study of kaonic deuterium X-rays by the SIDDHARTA experiment at DA$\Phi$NE }

\author[LNF]{M.~Bazzi}
\author[UVCA]{G.~Beer}
\author[SMI,LNF]{C.~Berucci}
\author[POLI]{L.~Bombelli}
\author[LNF,IFIN]{A.M.~Bragadireanu}
\author[SMI]{M.~Cargnelli\corref{cor}}
\ead{michael.cargnelli@oaaw.ac.at}
\author[LNF]{C.~Curceanu (Petrascu)}
\author[LNF]{A.~d'Uffizi}
\author[POLI]{C.~Fiorini}
\author[POLI]{T.~Frizzi}
\author[ROMA]{F.~Ghio}
\author[LNF]{C.~Guaraldo}
\author[TKY]{R.~Hayano}
\author[LNF]{M.~Iliescu}
\author[SMI]{T.~Ishiwatari}
\author[RIKEN]{M.~Iwasaki}
\author[SMI,MUN]{P.~Kienle\symbolfootnote[2]{deceased}}
\author[LNF]{P.~Levi Sandri}
\author[POLI]{A.~Longoni}
\author[SMI]{J.~Marton}
\author[RIKEN]{S.~Okada}
\author[LNF,IFIN]{D.~Pietreanu}
\author[IFIN]{T.~Ponta}
\author[SANTIAGO]{A.~Romero Vidal}
\author[LNF]{E.~Sbardella}
\author[LNF]{A.~Scordo}
\author[TKY]{H.~Shi}
\author[LNF,IFIN]{D.L.~Sirghi}
\author[LNF,IFIN]{F.~Sirghi}
\author[LNF]{H.~Tatsuno}
\author[IFIN]{A.~Tudorache}
\author[IFIN]{V.~Tudorache}
\author[MUN]{O.~Vazquez Doce}
\author[SMI]{E.~Widmann}
\author[SMI]{J.~Zmeskal}
\cortext[cor]{Corresponding author.}

\address[LNF]{INFN, Laboratori Nazionali di Frascati, C.P. 13, Via E. Fermi 40, I-00044 Frascati (Roma), Italy}
\address[UVCA]{Dep. of Physics and Astronomy, University of Victoria, P.O.Box 3055, Victoria B.C. Canada V8W3P6}
\address[SMI]{Stefan-Meyer-Institut f\"{u}r subatomare Physik, Boltzmanngasse 3, 1090 Wien, Austria}
\address[POLI]{Politecnico di Milano, Dip. di Elettronica e Informazione, Piazza L. da Vinci 32, I-20133 Milano, Italy}
\address[IFIN]{IFIN-HH, Institutul National pentru Fizica si Inginerie Nucleara Horia Hulubei, Reactorului 30, Magurele, Romania}
\address[ROMA]{INFN Sez. di Roma I and Instituto Superiore di Sanita I-00161, Roma, Italy}
\address[TKY]{University of Tokyo,7-3-1, Hongo, Bunkyo-ku,Tokyo, Japan}
\address[RIKEN]{RIKEN, Institute of Physical and Chemical Research, 2-1 Hirosawa, Wako, Saitama 351-0198 Japan}
\address[MUN]{Excellence Cluster Universe, Tech. Univ. M\"{u}nchen, Boltzmannstra{\ss}e 2, D-85748 Garching, Germany}
\address[SANTIAGO]{Universidade de Santiago de Compostela, Casas Reais 8, 15782 Santiago de Compostela, Spain}

\begin{abstract}

The study of the $\overline{K}N$ system at very low energies plays a key role for the understanding
of the strong interaction between hadrons in the strangeness sector. At the DA$\Phi$NE electron-
positron collider of Laboratori Nazionali di Frascati we studied kaonic atoms with Z=1 and Z=2,
taking advantage of the low-energy charged kaons from $\Phi$-mesons decaying nearly at rest. The
SIDDHARTA experiment used X-ray spectroscopy of the kaonic atoms to determine the transition yields and the strong interaction induced shift and width of the lowest experimentally accessible level (1s for H and D and 2p for He). Shift and width are connected to the real and imaginary part of the scattering length. To disentangle the isospin dependent scattering lengths of the antikaon-nucleon interaction, measurements of $K^-$p and of $K^-$d are needed. We report here on an exploratory deuterium measurement, from which a limit for the yield of the K-series transitions was derived: Y($K_{tot}$) $<$  0.0143 and Y($K_\alpha$) $<$  0.0039 (CL 90\%). Also, the upcoming SIDDHARTA-2 kaonic deuterium experiment is introduced.

\end{abstract}


\end{frontmatter}


\section{Introduction}
\label{intro}

The strong interaction between hadrons at low energies e.g., between the antikaon and the
nucleon ($\overline{K}N$) cannot be described in terms of quarks and gluons. Instead, effective field
theories are used which rely on experimental input. The data in the strangeness sector come from
kaon scattering experiments, atomic X-ray measurements and the energies, widths and branching ratios of known resonances.

The kaonic atom experiments extract the effect of the strong interaction on the low-lying
states by measuring X-ray transitions to these levels, seen as a shift and a width with respect
to the QED calculated values. These quantities are key inputs for theory and allow tests of the
ability of the theoretical models to accommodate the different types of experimental information.
For current theories see \cite{Meiszner2011,Weise2011} and references therein.

The $\overline{K}N$ interaction at rest is normally described in terms of complex scattering lengths.
To extract the isospin (I=0,1) dependent  $\overline{K}N$ scattering lengths,  $a_0$ and $a_1$, measurements of
kaonic deuterium are necessary along with kaonic hydrogen data. Equations (1) to (4) give the
connections between the proton, neutron and deuteron scattering lengths, $a_{K^-p}$, $a_{K^-n}$ and $a_{K^-d}$, and $a_0$ and $a_1$ where $m_{N}$ and $m_{K}$ denote the mass of the nucleon and of the kaon, C is a correction term accomodating nuclear effects.
\begin{eqnarray}
  a_{K^{-}p} &=& \frac{1}{2}[a_{0}+a_{1}] \\
  a_{K^{-}n} &=& a_{1} \\
  a_{K^{-}d} &=& \frac{4[m_{N}+m_{K}]}{[2m_{N}+m_{K}]}Q + C \\
  Q &=& \frac{1}{2}[a_{K^{-}p}+a_{K^{-}n}]=\frac{1}{4}[a_{0}+3a_{1}]
\end{eqnarray}

Kaonic atoms are systems bound electromagnetically, where the energy levels with strong
interaction ``switched off'' are known precisely from QED. The strong interaction modifies the
binding and causes the absorption of the kaon by the nucleus. This leads to a significant shift
and broadening of the $1s$ state of kaonic Z=1 atoms; higher states are only marginally affected. By
measuring the K-series X-ray transitions, the strong-interaction induced shift $\epsilon_{\rm{1s}}$ and width $\Gamma_{\rm{1s}}$ of
the ground state are directly observable. Via Deser-type formulae \cite{Deser-type} (see eq.5,\footnote{here the sign-reversed definition of the shift is used}  where $\alpha$ is the fine structure constant and $\mu_c$ the reduced mass of Kd), the shift and width are connected with the real and imaginary parts of the scattering length.
\begin{equation}
\epsilon_{\rm{1s}} - \frac{i}{2} \Gamma_{\rm{1s}}
= - 2 \alpha^3 \mu_c^2 a_{K^{-}d} \left(1 - 2 \alpha \mu_c
\left( {\rm ln} \alpha - 1 \right) a_{K^{-}d} \right)
\label{eq:1}
\end{equation}

A similar formula can be written for $a_{K^-p}$.

On the theoretical side there are many recent publications, giving consistent values of the
expected shift and width values, reported in table 1.

\begin{table} [ht]
\caption{Compilation of predicted $K^-$d scattering lengths \textit{$a_{K^{-}d}$}  and corresponding experimental quantities $\epsilon$$_{1s}$ and  $\Gamma$$_{1s}$ calculated from eq. 5 (except for \cite{Shevchenko2012} where the shift and width are given in the paper explicitly. They differ slightly for ``one-pole'' and ``two-pole'' structure of $\Lambda(1405)$, an averaged value is inserted in this table).  For fitting our data we use -805 eV for the shift and 750 eV for the width as representative values.}
 \label{table:4}
 \begin{center}
  \begin{tabular}{cccc} \hline
  \textit{$a_{K^{-}d}$} [fm] & $\epsilon$$_{1s}$ [eV] &  $\Gamma$$_{1s}$ [eV] & ref.\\
   \hline
   -1.58 + \textit{i} 1.37 & -887 & 757 & Mizutani 2013 \cite{Mizutani2013} \\
   -1.48 + \textit{i} 1.22 & -787 & 1011 & Shevchenko 2012 \cite{Shevchenko2012} \\
   -1.46 + \textit{i} 1.08 & -779 & 650  & Mei{\ss}ner 2011 \cite{Meiszner2011} \\
   -1.42 + \textit{i} 1.09 & -769 & 674  & Gal 2007 \cite{Gal2007} \\
   -1.66 + \textit{i} 1.28 & -884 & 665  & Mei{\ss}ner 2006 \cite{Meiszner2006} \\
   \hline
  \end{tabular}
 \end{center}
\end{table}

The kaonic hydrogen and deuterium X-ray measurements offer a unique possibility to determine the isospin dependent $\bar{K}N$ scattering lengths at threshold which are important inputs for theoretical studies of chiral SU(3) dynamics in low-energy quantum chromodynamics (QCD) and of the explicit chiral symmetry breaking due to the relatively large strange quark mass.

The SIDDHARTA (Silicon Drift Detectors for Hadronic Atom Research by Timing Application) experimental group performed the most precise measurement on kaonic hydrogen \cite{SIDKH} \cite{SIDKH-NPA}, the first measurement on kaonic helium 3 \cite{SIDKHe3} \cite{SIDKHe-width} and on gaseous helium 4 \cite{SIDKHe4} \cite{SIDKHe-width} and the first exploratory study of kaonic deuterium. The results for kaonic deuterium are reported in this paper.

\section{The SIDDHARTA Experiment}

Figure 1 illustrates the principle of the SIDDHARTA experiment. Charged kaon pairs delivered by the $\Phi$ decay pass through a trigger system and one of the kaons arrives in the deuterium target cell. Kaonic deuterium atoms are formed when negative kaons ($K^-$) enter the cell and loose their remaining kinetic energy through ionization and excitation and, by replacing an electron, are captured in an excited orbit (n $\simeq$ 25). Via different cascade processes,
the kaonic atoms deexcite to lower states. When the $K^-$  reach  low-n states with small angular momentum, they are absorbed by the nucleus. A small fraction of these kaonic deuterium atoms survive long enough to
make one of the (np-1s) transitions (K-series). The X-rays of these transitions are detected by
Silicon Drift Detectors (SDDs) placed around the target. Hits in the X-ray detectors coincident
with the trigger contain kaonic X-rays along with background from particles generated by kaon
decay and $K^-$ absorption. Due to the limited timing resolution some accidental (not kaon-correlated) background also remains.

\begin{figure} [ht]
 \begin{center}

\includegraphics*[scale=0.25]{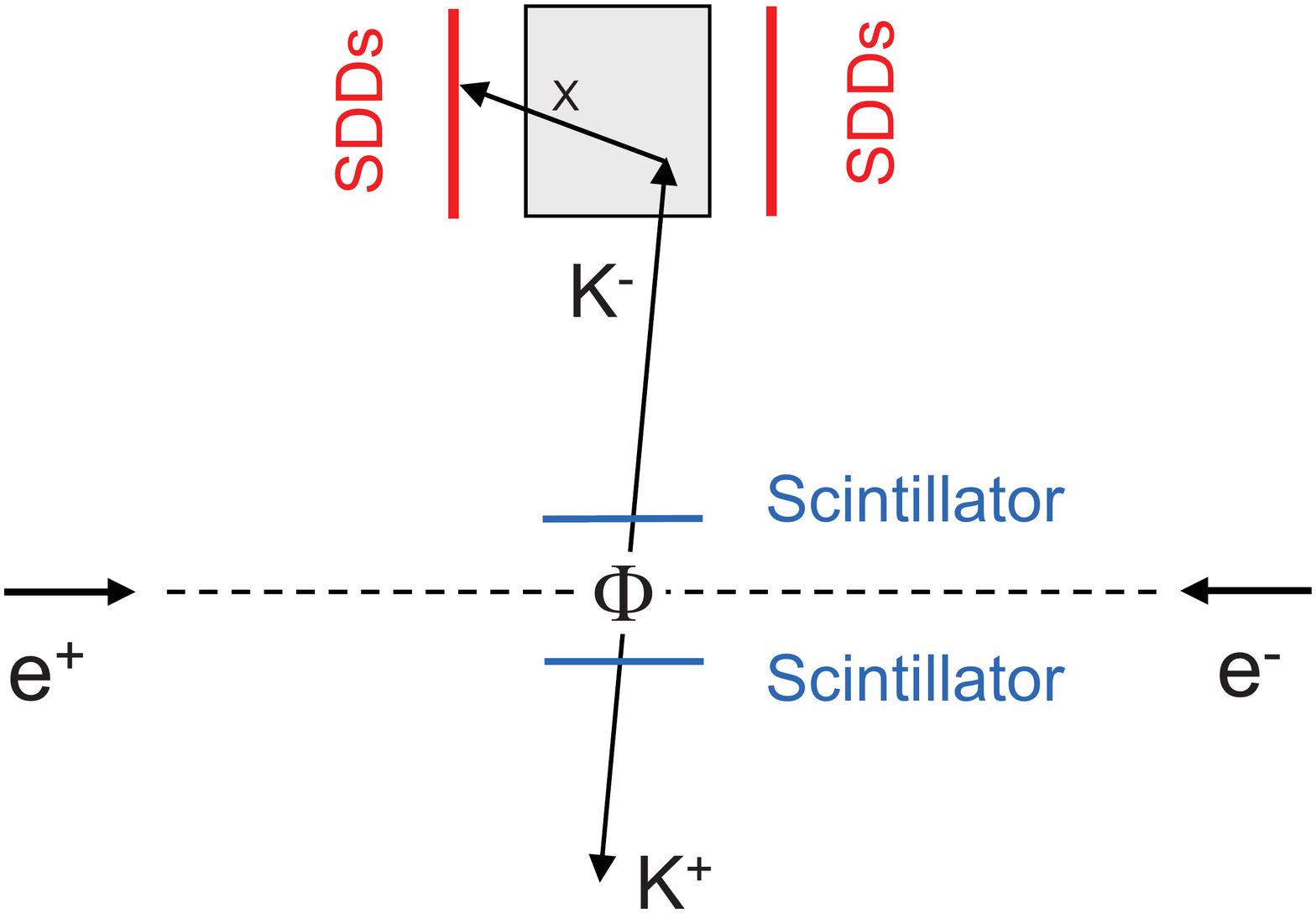}
  \caption{Sketch of the SIDDHARTA setup and of the experimental method.}
  \label{fig:Setup}
 \end{center}
\end{figure}

The DA$\Phi$NE $e^+$ $e^-$ collider  of Laboratori Nazionali di Frascati produced a peak luminosity of 3 $\times$ 10$^{32}$ cm$^{-2}$ s$^{-1}$ and an integrated usable\footnote{During injection of $e^{+}$ and $e^-$ into the rings, high beam-background prohibited X-ray measurements} luminosity of up to 8 pb$^{-1}$ per day.
The 2.2 liter cryogenic gas target was operated at 2.5 mg/cm$^{3}$, corresponding to 1.5 \% of liquid deuterium density (13.9 times STP density). The entrance window was located 16.8 cm above the interaction point (IP). The kaon trigger counter consisted of two 1.5 mm thick plastic scintillators situated 6 cm above and below the IP, read out by photomultipliers.




The experimental challenge is the very small kaonic deuterium X-ray yield and the difficulty in doing X-ray spectroscopy in the high radiation environment of an accelerator. Therefore new X-ray detectors \cite{NIM-tomo}\cite{sdd}\cite{sdd1} were developed and built in the framework of the SIDDHARTA project.






For each event, the data acquisition system stored  the signal amplitudes of the hits on the 144 SDDs and, if a kaon trigger was present, the time correlation between X-ray and kaon.


The basic energy calibration for each of the SDDs was obtained from periodic measurements of fluorescence lines from Ti and Cu foils excited by an X-ray tube with the beam on.
This procedure
delivered the information to compensate the individual gains, making it possible to sum the 144 energy spectra.
The response function of the summed detectors was found to deviate slightly from a pure Gaussian shape. Various approximations to account for this effect were tested, e.g., 2 exponential tail-functions convoluted in each fit evaluation step. More details on the calibration procedure are given in \cite{SIDKH}.



The kaonic deuterium data were collected in October 2009, for a total integrated luminosity of about 100 pb$^{-1}$.

\section{Data analysis and yield extraction}

The X-ray spectrum from the kaonic deuterium experiment is shown in Fig.2. No evident
signal corresponding to K-series X-rays is present. The data analysis, having the aim to extract
the yield of the K-transitions, proceeds as follows.

For the fit, the kaonic deuterium K-series transitions were put in the model with the pattern
of the transition energies taken from the electromagnetic values (E$_{2p-1s}$=7820 eV, E$_{3p-1s}$=9266 eV, E$_{4p-1s}$=9772 eV, E$_{5p-1s}$=10006 eV etc.\cite{Tomo1}), the shift set to -805 eV, the width to 750 eV to
represent the theory predictions (See table 1). For the shape of the transitions, Voigt functions were used, with the (energy dependent) Gaussian sigma fixed to the values obtained by interpolating the values from the calibration lines. The background lines from kaonic X-rays due to stops in the foils of the target window were Gaussians, the continuous background - a quadratic polynomial.
The relative intensities of the transitions were approximated to be the same as in the kaonic hydrogen measurement \cite{SIDKH}. The result of the fit is shown in Fig.2.

The poor statistics of the data does not allow fitting of the shift and width, the presence of the signal itself having a significance of 1.7 sigma. Including a possible signal into the fit model improves the chisquare from 143.3 for 162 data points and 137 degrees of freedom to 136.0 for 162 data points and 136 degrees of freedom.

\begin{figure} [ht]
\centerline{\includegraphics [scale=0.43 ] {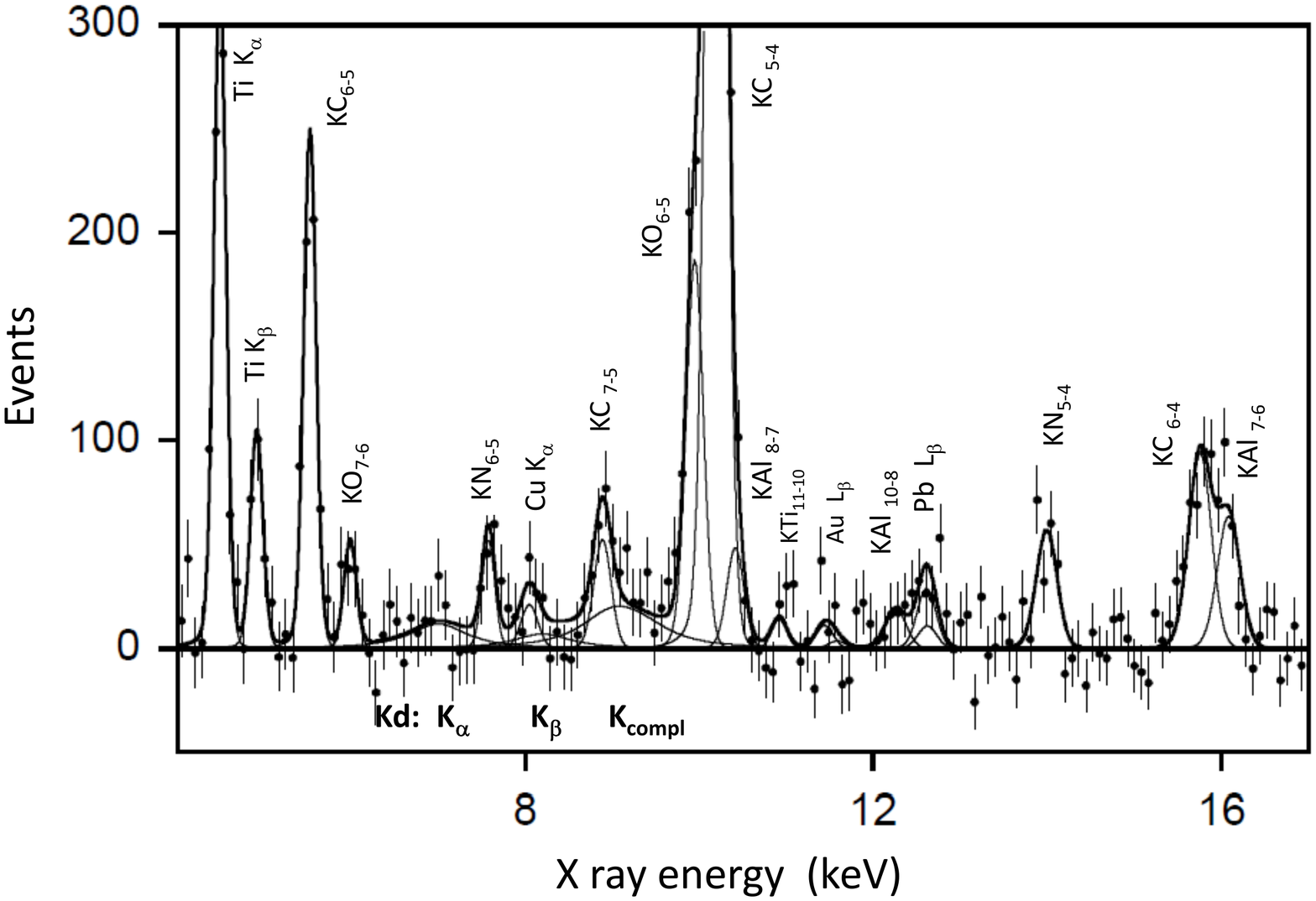}}
  \caption{X-ray spectrum from the kaonic deuterium experiment. The continuous background fit-component is already subtracted. Fit with fixed Kd transition shift and width (-805, 750 eV) and fixed yield ratio of the individual K-transitions. Integrated luminosity 100 pb$^{-1}$. The lines from kaonic X-rays due to stops in the window foils and from X-ray fluorescence excited by background are labeled. Note the excess of events in the region of a possible signal.}
 \label{fig:}
\end{figure}

To extract the yields of the kaonic deuterium transitions, we need to calculate the experimental efficiencies. We simulated the chain of processes starting from the produced $\Phi$ meson. Two independent studies were done using  GEANT3 and GEANT4 routines. The $\Phi$ mesons are produced in an interaction region with a spatial distribution defined by $\sigma$(horizontal), $\sigma$(vertical) and $\sigma$(beam). The momentum is given by the crossing angle of electrons and positrons which results in a horizontal boost of 25.5 MeV/c. The decay to charged kaon pairs is adapted to a sin$^2$($\theta$) distribution which is the result of a spin=1 particle decaying into two spin zero particles.

The charged kaons are followed along their tracks through the various materials (beam pipe, air, scintillators of the kaon trigger, degrader, vacuum window, target window and finally the target gas). When a negative kaon is stopped in the target or in the setup materials, a routine starts the tracks corresponding to the kaon-absorption products. For the case of deuterium, these are $\Sigma$s or $\Lambda$s, pions and nucleons. The kaonic X-rays were generated isotropically with a yield of 100 percent. Following these X-rays, the attenuation is treated by a user routine and finally the eventual absorption in the SDDs is registered.

The relevant number to derive an absolute kaonic deuterium X-ray yield from the experimental data is the calculated efficiency-per-trigger, W, defined as ``Kd K-series X-rays detected in the SDDs'' per ``kaon pair detected in the kaon-trigger'' (for 100 \% X-ray yield and data from 100 SDDs). The contributions to the systematic error are shown in Table 2, a compilation of materials in Table 3.

\begin{table}
\caption{Sources of systematic errors of the calculated efficiency.  }
 \label{table:1}
 \begin{center}
  \begin{tabular}{ccc} \hline \hline
  error source (uncertainty)  & contribution to efficiency error\\
\hline
total material thickness ($\pm$ 7 mg/cm$^2$) & $\pm$ 9 \% \\
size of  $e^+$ $e^-$ collision zone ($\pm$ 5 mm) & $\pm$ 8 \%  & \\
setup geometry, distance from IP ($\pm$  5 mm) & $\pm$ 7 \%  \\
variation in  $e^+$ $e^-$ crossing angle ($\pm$ 2 mrad) & $\pm$ 4 \%  \\
gas density ($\pm$ 2 \%)  & $\pm$ 3 \%  \\
statistical error Monte Carlo & $\pm$ 3 \% \\
central value of beam tune ($\pm$ 0.4 MeV) & $\pm$ 4\%  &  \\
\hline
quadratic sum of above & $\pm$  17 \% \\

   \hline
  \end{tabular}
 \end{center}
\end{table}

We obtained, after averaging the 2 Monte Carlo efficiency calculations, and using the larger errors for W:

W = 0.00752  with a relative error of  17 \% \\


The fit of the spectrum gave a total number of kaonic deuterium events of

 $N_X$ = 518 $\pm$ 250 (stat.) $\pm$ 180 (syst.) \\

Adding statistical and systematic error quadratically we obtain
$N_X$ = 518 $\pm$ 308  corresponding to a significance of 1.7 sigma.

The systematic error of $N_X$ covers the dependence of the resultant intensity on the selection of the fit range and the background function, the channel binning, the choice of detectors, the selected correlation time-window and on variations of the signal yield pattern in the limits given from the Kp measurement and on the signal shift and width which were varied by $\pm$ 60 eV. \\

Putting together the above number with the number of detected kaon pairs, the efficiency and correction factors (e.g., deadtime), we obtain, after adding error components quadratically and using the larger of the asymmetric errors of 1/W: \\

total Kd K-series yield:   Y($K_{tot}$) =  0.0077 $\pm$  0.0051

This corresponds to a Kd $K_\alpha$ yield Y($K_\alpha$) = 0.0019 $\pm$  0.0012 \\

Such values are compatible with what is expected, namely a yield of a factor about 10 smaller then the KH yield, estimated to be 1 to 2 $\%$ for $K_\alpha$. See \cite{casc-faifman-ivanov,casc-jensen,casc-koike,casc-raeisi-kalantari} for predicted yields and their dependence on the target density and on the strong interaction parameters (the 1s shift, the 1s width and the width of the 2p state  $\Gamma_{2p}$). The larger absorption in the 2p state (larger $\Gamma_{2p}$) in the case of kaonic deuterium atoms as compared to kaonic hydrogen, causes the lower $K_\alpha$ yield.

Note that the ratio Y($K_\alpha$) / Y($K_{tot}$) was fixed in the fit function,  assuming the value to be the same as the one obtained by SIDDHARTA for kaonic hydrogen.\\

From the yield values given above, the upper limits for the yields are (CL 90\%):\\
Y($K_{tot}$) $<$ 0.0143 \\
Y($K_\alpha$) $<$ 0.0039  \\

\begin{table}
\caption{Compilation of setup materials which the kaons are passing. Due to the $\Phi$ boost the kinetic energy of kaons is direction dependent, this effect is compensated by a shaped degrader consisting of stripes of 2 cm wide foils of varying thickness.  }
 \label{table:1}
 \begin{center}
  \begin{tabular}{ccc} \hline \hline
description  & thickness in mm & thickness in  mg/cm$^2$ \\
\hline
beam pipe window (Al) & 0.400 & 108.0 \\
air & 141 & 17.0\\
plastic scintillator & 1.450 & 150.0 \\
scint. wrapping (tape) & 0.500 & 71.0 \\
scint. wrapping (Al) & 0.006 & 1.62 \\
uniform degrader (Mylar) & 0.175 & 24.3 \\
shaped degrader (Mylar) & 0.200 - 0.700 & 27.8 - 93.7 \\
light shield (paper) & 0.120 & 96.0 \\
vacuum window (Kapton) & 0.125 & 177.5 \\
calibration foil (Ti) & 0.025 & 11.3 \\
target window (Kapton) & 0.100 & 14.2 \\
\hline
sum of above & & 698.7 - 764.6 \\
   \hline
  \end{tabular}
 \end{center}
\end{table}

\section{The SIDDHARTA-2 experiment}

A quantitative measurement of kaonic deuterium X-rays requires a substantial improvement of the applied experimental technique. Since the yield is expected to be about 1/10 of the KH yield and the width up to twice the KH width, we need a reduction in background of a factor about 20 and an increase in the  detection efficiency of 2-3. For the simulations we use a yield Y($K_\alpha$) = 0.001 which is a conservative assumption given the predictions from theory \cite{casc-faifman-ivanov,casc-jensen,casc-koike,casc-raeisi-kalantari}. The reliability of these predictions can be estimated to be better then to a factor of 2, derived from our experiences from kaonic hydrogen (to be published).

\begin{figure} [ht]
\centerline{\includegraphics [scale=0.3] {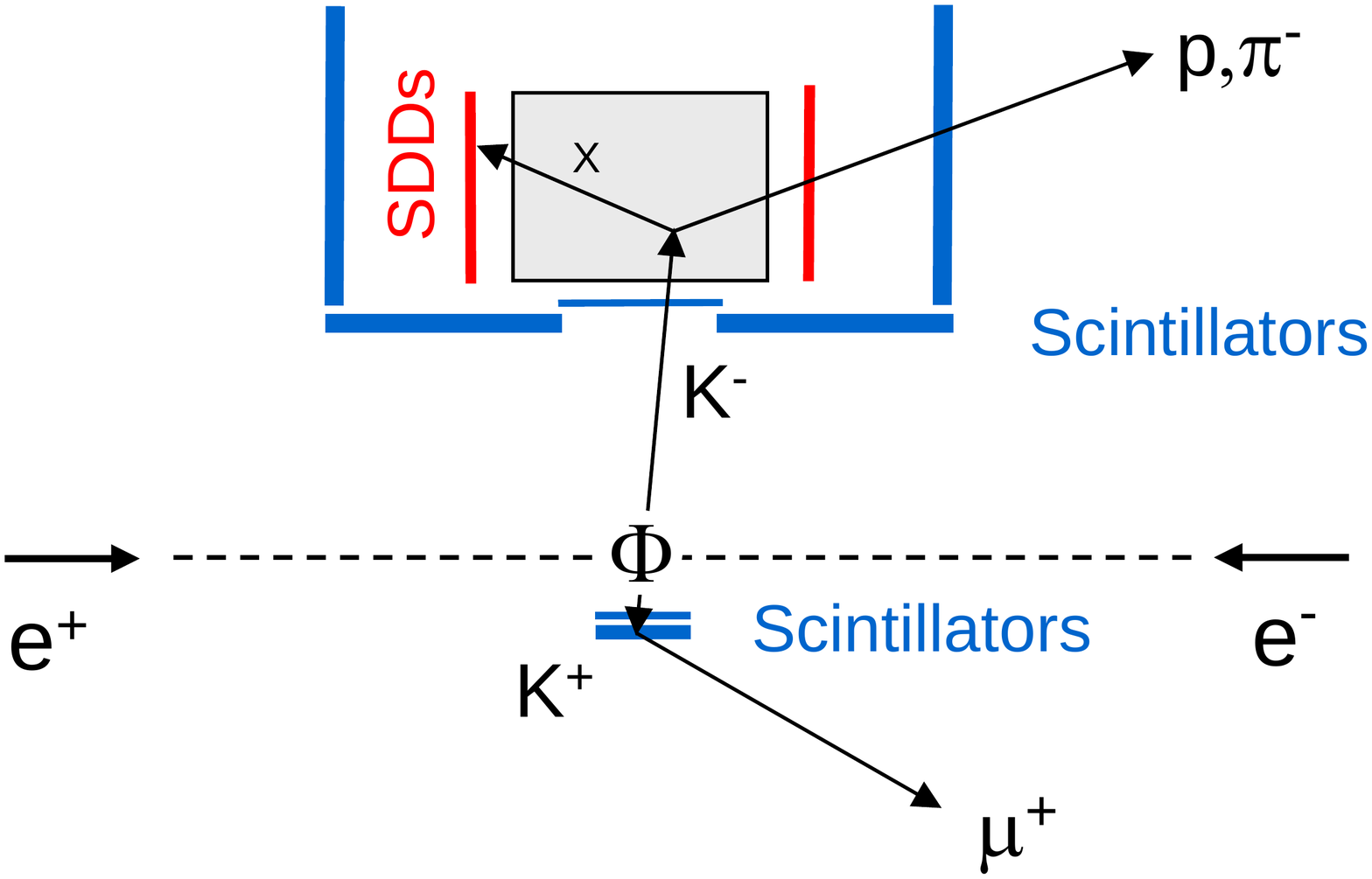}}
  \caption{Sketch of the SIDDHARTA-2 setup and experimental method. As compared to Fig.1 note the changed position of the upper trigger scintillator and the additional scintillators used to detect kaon secondaries and beam-background.}
 \label{fig:}
\end{figure}

The upgrade of SIDDHARTA to SIDDHARTA-2 is based on four main modifications; shown in Fig.3:

(1) Trigger geometry and target density: By placing the upper kaon-trigger detector close in
front of the target entrance window, the probability that a triggered kaon really enters the gas and
is stopped there is much improved. Making the detector smaller than the entry area gives away
some signal, but suppresses efficiently the kaonic lines from ``wallstops'' (kaons entering the gas
volume, but passing from the inside of the target to the cylindrical walls). The number ``signal per trigger'' goes up, which also reduces the accidental background coming along with every trigger. We plan as well to double the gas density which enhances the gas stops and further reduces the wall-stops.

(2) $K^+$ discrimination to suppress kaon decay background: A ``kaon stopper'' scintillator is
placed directly below the lower kaon trigger scintillator. When a $K^-$ is stopped there, only one
(large) signal from pileup of stopping and kaon-absorption secondaries is seen, whereas when
a $K^+$ is stopped, the kaon-decay particles are seen after the signal from the stopping (mean $K^+$
lifetime 12.8 ns). Using a flash-ADC we will be able to efficiently distinguish the 2 cases. In
addition, we will use scintillators surrounding the target to measure $K^-$ absorption secondaries
(pions). The time window for gas stops is about 4 ns wide. By this condition we also suppress
stops in the entry window.

(3) Active shielding: The scintillators surrounding the target will also be used in prompt anticoincidence if the spatial correlation of SDD and scintillator hits indicates that it originated from a pion (``charged particle veto''). An anticoincidence covering the SDD time window of about 600 ns (with the exception of the 4 ns of the gas stopping time) will reduce the accidental background. Although the scintillators have only low efficiency for gammas, the abundance of secondaries from the electromagnetic showers allows a relevant reduction of accidental (``beam'') background. The upper trigger scintillator has 2 functions, it is also used as an anticoincidence counter: after the kaon and eventual prompt kaon-absorption secondaries pass, it vetos beam-background.

(4) Operating SDDs at a lower temperature: tests indicate that an improvement of the timing resolution by a factor of 1.5 is feasible by more cooling.

\begin{figure} [ht]
\centerline{\includegraphics [scale=0.42] {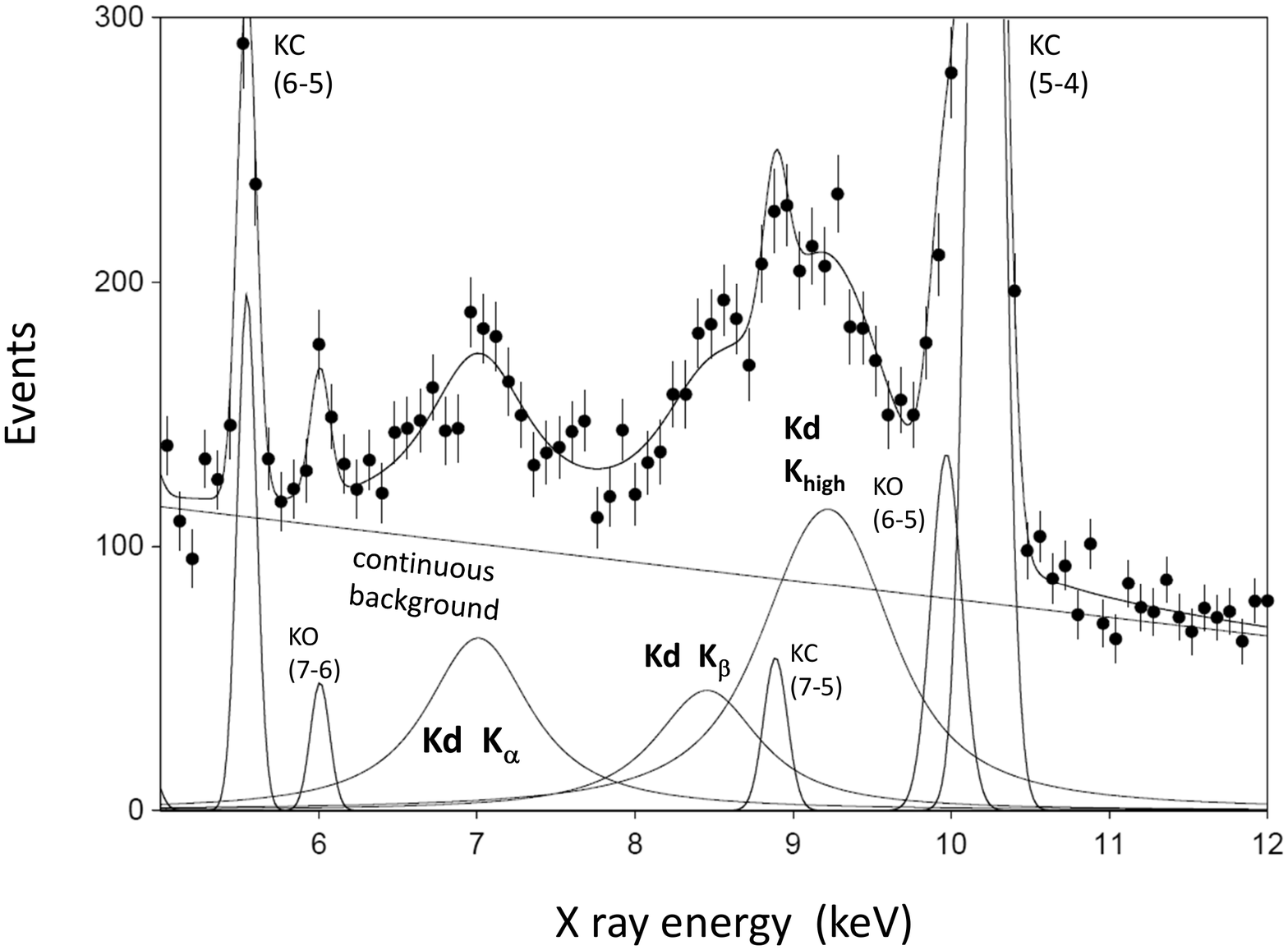}}
  \caption{Monte Carlo simulation of the kaonic deuterium spectrum corresponding to the proposed SIDDHARTA-2 techniques. Shift and width values: -805 and 750 eV respectively. Yield Y($K_\alpha$) = 0.001. Integrated luminosity 800 pb$^{-1}$.  }
 \label{fig:}
\end{figure}

The signal enhancement by a factor 2 to 3 is due to moving the target cell closer to the IP, by changing its shape, by a better solid angle of the SDDs and by the higher gas density.

In such conditions, with an integrated luminosity of 800 pb$^{-1}$ a precision of about 70 eV for the shift, and 160 eV for the width are attainable, resulting in a relative precision similar to that obtained for kaonic hydrogen.

\section{Summary and conclusions}

The SIDDHARTA experiment has resulted in unprecedented measurements of light kaonic atoms. The kaonic deuterium exploratory study performed by SIDDHARTA for the first time, delivered a 1.7 sigma hint of a signal if shift and width of the K-series transitions are taken from theory. The experimental total Kd K-series yield was obtained to be Y($K_{tot}$) = 0.0077 $\pm$  0.0051 corresponding to a Kd $K_{\alpha}$ yield of Y($K_\alpha$) = 0.0019 $\pm$  0.0012. In upper limits this means Y($K_{tot}$) $<$  0.0143 and Y($K_\alpha$) $<$  0.0039  (CL 90\%).

Based on these results, we worked out a scheme for an improved experimental technique ``SIDDHARTA-2'' which promises to deliver a sound quantitative measurement of the shift and width in kaonic deuterium, see Fig.4, long awaited for the description of the strong interaction at lowest energies in hadronic systems with strangeness.

\section*{Acknowledgement}

This paper is dedicated to the memory of Paul Kienle, distinguished member of the SIDDHARTA Collaboration, who played an eminent role in many sectors of experimental modern physics, including  the
field of strong interaction with strangeness.

We thank C. Capoccia, G. Corradi, B. Dulach, and D. Tagnani from LNF-INFN; and H. Schneider, L. Stohwasser, and D. St\"uckler from Stefan-Meyer-Institut, for
their fundamental contribution in designing and building the SIDDHARTA
setup. We thank as well the DA$\Phi$NE staff for the excellent working conditions
and permanent support. Part of this work was supported by the European Community-Research
Infrastructure Integrating Activity ``Study of Strongly Interacting Matter''
(HadronPhysics2, Grant Agreement No. 227431, and HadronPhysics3 (HP3) Contract No. 283286) under the Seventh
Framework Programme of EU;
HadronPhysics I3 FP6 European Community program, Contract No. RII3-CT-2004-506078;
Austrian Science Fund (FWF) [P24756-N20];
Austrian Federal Ministry of Science and Research BMBWK 650962/0001 VI/2/2009;
Romanian National Authority for Scientific Research, Contract No. 2-CeX 06-11-11/2006;
and the Grant-in-Aid for Specially Promoted Research (20002003), MEXT, Japan.


\begin{thebibliography}{00}

\bibitem{Meiszner2011} M.~D\"oring, U.-G.~Mei{\ss}ner, Phys. Lett. B 704 663 (2011).
\bibitem{Weise2011} Y.~Ikeda, T.~Hyodo, W.~Weise, Phys. Lett. B 706 63 (2011).
\bibitem{Deser-type} U.-G.~Mei{\ss}ner, U.~Raha, A.~Rusetsky, Eur. Phys. J. C 35 349 (2004).
\bibitem{Mizutani2013} T.~Mizutani, C.~Fayard, B.~Saghai, K.~Tsushima, arXiv:1211.5824[hep-ph] (2013).
\bibitem{Shevchenko2012} N.V.~Shevchenko, Nucl. Phys. A 890-891 (2012) 50-61.
\bibitem{Gal2007} A. Gal,  Int. J. Mod. Phys. A22 (2007) 226.
\bibitem{Meiszner2006} U.-G. Mei{\ss}ner, U. Raha, and A. Rusetsky, Eur. Phys. J  C47 (2006) 473.
\bibitem{SIDKH} M.~Bazzi, {\em et al.} (SIDDHARTA Collaboration),
	 Phys. Lett. B 704 (2011) 113.
\bibitem{SIDKH-NPA} M.~Bazzi, {\em et al.} (SIDDHARTA Collaboration),
	 Nucl. Phys. A 881 (2012) 88.
\bibitem{SIDKHe3} M.~Bazzi, {\em et al.} (SIDDHARTA Collaboration),
	 Phys. Lett. B 697 (2011) 199.
\bibitem{SIDKHe4} M.~Bazzi, {\em et al.} (SIDDHARTA Collaboration),
	 Phys. Lett. B 681 (2009) 310.
\bibitem{SIDKHe-width} M.~Bazzi, {\em et al.} (SIDDHARTA Collaboration),
	 Phys. Lett. B 714 (2012) 40.
\bibitem{NIM-tomo} M. Bazzi, {\it et al.}, Nucl. Instr. and Meth. A 628, (2011) 264.
\bibitem{sdd} C. Fiorini, {\it et al.}, Nucl. Instr. and Meth. A 568 (2006) 322.
\bibitem{sdd1} P. Lechner, {\it et al.}, Nucl. Instr. and Meth. A 458 (2001) 281.
\bibitem{Tomo1} T. Ishiwatari, SMI internal report (2007).
\bibitem{casc-faifman-ivanov} M. Faber {\it et al.}, Phys. Rev. C 84, 064314 (2011).
\bibitem{casc-jensen} T.S. Jensen, Frascati Physics Series Vol. XXXVI (2004).
\bibitem{casc-koike} T. Koike, {\it et al.}, Phys. Rev. C 53, 79-87 (1996).
\bibitem{casc-raeisi-kalantari} M. Raeisi G. and S. Z. Kalantari, Phys. Rev. A 79, 012510 (2009).




\end{thebibliography}
\end{document}